\documentclass[iop]{emulateapj}
\usepackage{natbib}
\usepackage{graphicx}
\usepackage{graphics}
\usepackage{amsmath}
\usepackage[usenames]{color}
\newcommand{\bit}[1]{\mbox{\textbf{\emph{#1}}}}

\newcommand       \be           {\begin{equation}}
\newcommand       \ee           {\end{equation}}

\newcommand{\jcap}{JCAP}

\def\l{\ell}
\def\d{\mathrm{d}}

\def\fsky{f_\mathrm{sky}}
\def\hmpc{h^{-1}\mathrm{Mpc}}
\def\alm{a_{\ell m}}
\def\pl{P_{\ell}}

\def\plhat{\hat{P}_{\ell}}
\def\pltil{\tilde{P}_{\ell}}
\def\plbar{\bar{P}_{\ell}}
\def\plund{P_{\ell}^\mathrm{u}}
\def\fnl{f_\mathrm{NL}}
\def\ng{n_\mathrm{g}}
\def\arcm{\mbox{arcmin}}
\def\Om{\Omega_\mathrm{m}}
\def\Ob{\Omega_\mathrm{b}}
\def\bg{b_\mathrm{g}}
\def\ns{n_\mathrm{s}}
\def\s8{\sigma_8}

\begin{document}

\title{Likelihood of the
Power Spectrum in Cosmological Parameter Estimation}

\author{Lei Sun, Qiao Wang and Hu Zhan}
\affil{Key Laboratory of Optical Astronomy, National Astronomical Observatories, 
Chinese Academy of Sciences, 20A Datun Road, Chaoyang District, Beijing 100012, China}
\email{sunl@nao.cas.cn}

\begin{abstract}
The likelihood function is a crucial element of
parameter estimation. In analyses of galaxy overdensities and weak
lensing shear, one often approximates the likelihood of the power
spectrum with a Gaussian distribution. The posterior probability
derived from such a likelihood deviates considerably from the exact 
posterior on the largest scales probed by any survey, where the 
central limit theorem does not apply. We show that 
various forms of Gaussian likelihoods can have a significant impact
on the estimation of the primordial non-Gaussianity parameter $\fnl$ 
from the galaxy angular power spectrum. The Gaussian plus log-normal 
likelihood, which has been applied successfully in analyses of the 
cosmic microwave background, outperforms the Gaussian likelihoods. 
Nevertheless, even if the exact likelihood of the power spectrum is used, 
the estimated parameters may be still biased. As such, the likelihoods 
and estimators need to be thoroughly examined for potential systematic 
errors.

\end{abstract}

\section{Introduction}
\label{sec:int}

Bayesian inference is widely practiced in cosmological parameter
estimation. The posterior distribution of the parameters (given the
observed data) is mapped from the product of the likelihood of the
data (given the parameters) and the prior of the parameters according
to Bayes' Theorem. The likelihood extracts the information from the
observed data, whereas the prior is from external sources and often
assumed to be flat for the lack of external knowledge.
Therefore, the likelihood function plays a crucial role in
parameter estimation and needs to be determined accurately.

Cosmic fluctuations are expected to be well described by a Gaussian
random field on scales where nonlinearity is negligible,
so that the Fourier modes of the fluctuations
follow independent complex Gaussian distributions characterized by
the power spectrum \citep[][referenced herein]{bardeen86,bond87}.
The observed power of fluctuations at a given scale then follow a
Gamma distribution that is also determined by the power spectrum.
For a Gaussian random field,
the power spectrum encapsulates all the information in a small
set of numbers and, hence, can be analyzed in place of the
fluctuations far more efficiently without loss of information
\citep{tegmark97}.

It is noted in cosmic microwave background (CMB) analyses that the
Gaussian approximation of the power spectrum likelihood leads
to parameter biases, and better approximations have been developed
\citep{bond2000, bartlett2000, verde03}.
In analyses of galaxy density
fluctuations and weak lensing shear fluctuations, however, the
Gaussian approximation remains the backbone of the standard practice,
and the covariance of the observables is often taken to be
independent of cosmology \citep[e.g.,][]{tegmark06, percival10, ho12,
hoekstra06, semboloni06, massey07, benjamin07}.
Although the central limit theorem
guarantees Gaussianity of the power spectrum likelihood on scales
much smaller than the dimensions of a survey, there is always
considerable deviation at the largest scales probed by the survey.

Recently, the model dependence of the covariance of the Gaussian
likelihood has drawn some attention. It is found to have a
significant impact on weak lensing shear analyses (\citealt{eifler09};
\citealt{jee13}; but cf. \citealt{kilbinger13}) and a mild
effect on baryon acoustic oscillations (BAO) analyses
\citep{labatie12}. Separately, \citet{wilking13} propose a 
quasi-Gaussian method by applying the Gaussian approximation on an 
unconstrained variable that is transformed from the constrained 
correlation functions \citep{Keitel11, Schneider09}. They find it a 
better approximation than the ususal Gaussian approximation.
Although these studies are all based on correlation functions, they
motivate a closer examination of the approximations in the
likelihood analyses of power spectra.
In fact, \citet{carron13} has shown based
on Fisher information that including the model dependent covariance
would underestimate the parameter uncertainties.

To demonstrate the effect of approximate likelihood functions, we
generate mock galaxy angular power spectra and estimate cosmological
parameters using several approximate likelihoods. Particular
attention is given to the primordial non-Gaussianity parameter
$f_\mathrm{NL}$, whose effect is most prominent on the largest scales
\citep{dalal08, matarrese08}.

This paper is organized as follows. In Section~\ref{sec:like}, we
introduce the likelihood of power spectrum and its four approximations.
We study the impact of the approximate likelihoods on parameter estimation in Section~\ref{sec:impact}.
Section~\ref{sec:sum} is a summary.

\section{Likelihood of the Power Spectrum}
\label{sec:like}

Bayes' theorem relates the posterior probability
$\mathcal{P}(\boldsymbol{\theta}|\bit{D})$ of the parameters
$\boldsymbol{\theta}$ given the data $\bit{D}$ to the likelihood
of the data $\mathcal{L}(\bit{D}|\boldsymbol{\theta})$ given
the parameters:
\be
{\cal P}(\boldsymbol{\theta}|\bit{D}) \propto
\mathcal{P}(\boldsymbol{\theta})
{\cal L}(\bit{D}|\boldsymbol{\theta}),
\label{eq:bayes}
\ee
where $\mathcal{P}(\boldsymbol{\theta})$ is the prior of the
parameters, and a normalization factor depending only on the data
has been dropped. With a flat prior, the task of parameter estimation
is essentially mapping $\mathcal{P}(\boldsymbol{\theta}|\bit{D})$
from $\mathcal{L}(\bit{D}|\boldsymbol{\theta})$. In this Letter,
our dataset is taken to be the galaxy angular power spectrum.

On scales where the cosmic density field can be treated as a Gaussian
random field, the real and imaginary parts of the spherical harmonic
coefficients $\alm$ of the density fluctuations both follow a
Gaussian distribution with zero mean and variance $\frac{1}{2} \pl$,
where $\pl \equiv \langle \vert\alm\vert^2\rangle$ is the angular
power spectrum at multipole $\l$. For an ideal full-sky survey without
measurement noise, the mean power of the modes
$\plhat=\frac{1}{2\l+1} \sum_m \left\vert\alm\right\vert^2$
(hereafter we refer to it less rigorously as the ``observed'' power 
spectrum) is an unbiased estimator of the angular power spectrum,
and it follows a Gamma distribution\footnote{The combined term
$(2\l+1)\plhat/\pl$ follows a $\chi^2$ distribution, a
special case of the Gamma distribution.} at each $\l$:
\be
\label{eq:L_Gamma}
\mathcal{L}_{\Gamma}(\plhat | \pl) \propto
\frac{1}{\pl} \left(\frac{\plhat}{\pl}\right)^{\frac{2\l+1}{2}-1}
\exp\left[-\frac{(2\l+1)\plhat}{2\pl}\right],
\ee
whose mean and variance are $\pl$ and $\frac{2}{2\l+1} \pl^2$,
respectively. One may roughly account for the effect of partial sky
coverage by replacing $2\l+1$ with $(2\l+1)\fsky$
\citep{scott94, hobson96}.
For galaxy surveys, the power spectra include a contribution from
shot noise $\ng^{-1}$, where $\ng$ is the surface number
density of galaxies.

With a single Gaussian random field one can express the likelihood
of the whole
power spectrum as a product of likelihoods of each multipole, i.e.,
$\mathcal{L}_\Gamma(\hat{\bit{P}}|\bit{P}) =
\prod_\l \mathcal{L}_{\Gamma}(\plhat | \pl)$.
In reality, significant correlations between different
multipoles can arise from various sources such as a sky cut.
Tomographic analyses of galaxies and shear in multiple photometric
redshift bins also need to account for the correlations between
different bins. Computing the exact likelihood in such
cases is not practical, so approximations are necessary
\citep[e.g.,][]{verde03, carron13}.

We consider the following approximate likelihoods: Gaussian
\be \label{eq:L_G}
\mathcal{L}_\mathrm{G}(\plhat \mid \pl) \propto
\sqrt{\frac{2\l + 1}{2 \pl^2}}
\exp\left[-\frac{(2\l + 1)(\plhat - \pl)^2}{4 \pl^2}\right];
\ee
Gaussian without the determinant
\be \label{eq:L_Gnd}
\mathcal{L}_\mathrm{G,nd}(\plhat \mid \pl) \propto
\exp\left[-\frac{(2\l + 1)(\plhat - \pl)^2}{4 \pl^2}\right];
\ee
and Gaussian plus log-normal \citep{verde03}
\be \label{eq:L_GLN}
\mathcal{L}_\mathrm{G+LN} \propto \mathcal{L}_\mathrm{G,nd}^{1/3}
\exp\left[-\frac{2\l + 1}{6}\ln^2\left(\frac{\plhat}{\pl}\right)
\right].
\ee
It is customary to replace $\frac{2}{2\l+1}\pl^2$ in the
exponent of Equation~(\ref{eq:L_Gnd}) with a constant variance
\citep[or constant covariance in multivariate case, e.g.,][]
{tegmark97}. We refer to such an approximation as $\mathcal{L}_\mathrm{G,cc}$.

\begin{figure}
\includegraphics[width=\columnwidth]{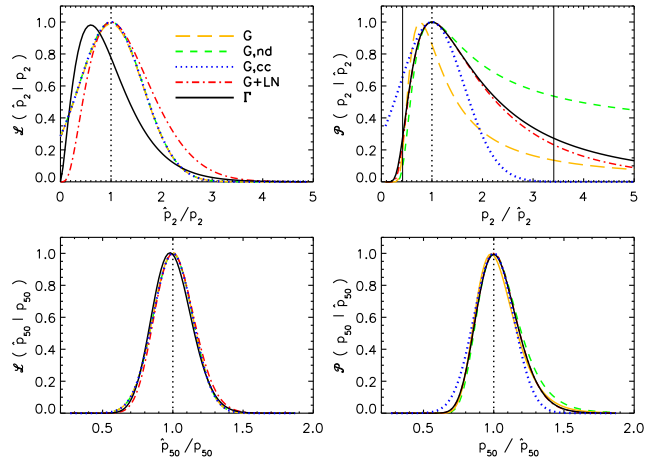}
 \caption{\emph{Upper left panel}: The true likelihood function of
the quadrupole $\mathcal{L}_{\Gamma}(\hat{P}_2 | P_2)$
(solid line) and its four approximations: Gaussian (long
dashed line), Gaussian without the determinant (short dashed line),
Gaussian with a constant covariance (dotted line) and
Gaussian plus log-normal (dot-dashed line).
\emph{Lower left panel}: Same as the upper left panel but for
$\l = 50$.
\emph{Upper right panel}: The posterior probabilities of the
underlying quadrupole $\mathcal{P}(P_2 | \hat{P}_2)$
mapped from the likelihood functions in the upper left panel.
The two vertical thin solid lines mark the 
1-$\sigma$ confidence interval $[0.42, 3.41]$ for
$\mathcal{P}_\Gamma$.
\emph{Lower right panel}: Same as the upper right panel but for
$\l = 50$. The likelihood functions and posterior probabilities
are scaled to have a maximum of unity.}
\label{fig:like-post}
\end{figure}

Figure~\ref{fig:like-post} shows the true likelihood function of
the angular power spectrum $\mathcal{L}_{\Gamma}(\plhat | \pl)$ and
its four approximations (left panels) along with corresponding
posterior probabilities $\mathcal{P}(\pl | \plhat)$ (right panels).
None of the approximate likelihoods matches
$\mathcal{L}_{\Gamma}$ for the quadrupole. However, from the
point of view of parameter estimation, it is most important to
reproduce the posterior probability accurately. Indeed, the upper
right panel demonstrates that the posterior mapped from the
Gaussian plus log-normal likelihood matches that from
the true likelihood fairly well.

The Gaussian approximation $\mathcal{L}_\mathrm{G}$ results in a
biased maximum-likelihood estimate of the quadrupole. Although its
posterior $\mathcal{P}_\mathrm{G}$ appears to underestimate the
uncertainty of the quadrupole, all moments of
$\mathcal{P}_\mathrm{G}$ diverge with increasing range of $\pl$
(see Section~\ref{sec:asm}).
The posterior probabilities of the quadrupole mapped from
$\mathcal{L}_\mathrm{G,nd}$ and $\mathcal{L}_\mathrm{G,cc}$
recover the correct maximum likelihood value $\pl=\plhat$, but
their shapes deviate significantly from $\mathcal{P}_\Gamma$.
In addition, $\mathcal{P}_\mathrm{G,nd}$ also suffers from
divergent moments.

For higher multipoles, the true likelihood function approaches
Gaussian because of the central limit theorem. The approximate
forms considered become less distinguishable from
$\mathcal{L}_{\Gamma}$.
Therefore, the difference between the posterior probabilities is
much smaller at larger $\l$s, which is illustrated in the lower
panels of Figure~\ref{fig:like-post} with $\l = 50$.

Although the Gaussian plus log-normal approximation of the power 
spectrum likelihood is fairly 
accurate even at $\ell = 2$, precision CMB analyses continue to
motivate efforts to improve the approximation or analyis method.
For example, in \emph{WMAP} 3-year results the large-scale modes
($\l \le 30$) are analyzed in pixel space with a Gaussian 
likelihood to improve the estimation of the power spectrum 
\citep{hinshaw07}. \citet{planckXV} adopt an approximation 
proposed by \citet{hamimeche08}, which reduces to the Gamma 
distribution in the ideal case considered in this paper.

\section{Impact of Approximate Likelihoods on Parameter Estimation}
\label{sec:impact}

In this section, we examine the performance of the approximate
likelihood functions in terms of parameter biases and
uncertainties. We first analyze the likelihoods and posteriors
for a single mode to identify potential issues and then estimate
parameters from mock galaxy angular power spectra to evaluate
these likelihoods in a more realistic way.

\subsection{Analyses with a Single Mode}
\label{sec:asm}

We assume for simplicity that the parameter of interest is a linear
function of the power spectrum $\pl$. In this case, one only needs
to be concerned with estimating $\pl$ from the observed $\plhat$.
Since $\plhat$ itself is an unbiased estimator of $\pl$ constructed
from $\alm$, an estimator that takes $\plhat$ as an input and
returns the value of $\plhat$ as the best estimate of $\pl$ would
also be unbiased. Hereafter we label the underlying power spectrum 
to be estimated as $\plund$ to distinguish it the from the generic
notation of a power spectrum.

From Equations~(\ref{eq:L_Gamma})--(\ref{eq:L_GLN}) and the right
column of Figure~\ref{fig:like-post}, one sees that the posteriors
$\mathcal{P}_\mathrm{G,nd}$, $\mathcal{P}_\mathrm{G,cc}$,
$\mathcal{P}_\mathrm{G+LN}$, and $\mathcal{P}_\Gamma$ all peak at
$\plhat$, so we consider the mode of these posteriors, $\pltil$,
as an estimator (also known as the maximum likelihood estimator).
In terms of the ensemble behavior,
\begin{align} \label{eq:pl_ml}
\langle \pltil \rangle &= \int \plhat
\mathcal{L}_{\Gamma}(\plhat|\plund) \d\plhat = \plund  \\
\sigma_{\pltil}^2 &= \int \plhat^2
\mathcal{L}_{\Gamma}(\plhat|\plund) \d\plhat -
\langle \pltil \rangle^2 = \frac{2}{2\l+1}\left(\plund\right)^2
\end{align}
as desired. Hence, $\pltil$ is a good estimator of $\plund$ for
$\mathcal{P}_\mathrm{G,nd}$, $\mathcal{P}_\mathrm{G,cc}$,
$\mathcal{P}_\mathrm{G+LN}$, and
$\mathcal{P}_\Gamma$. However, $\pltil$ is biased for
$\mathcal{P}_\mathrm{G}$ and behaves as
$(1-\l^{-1})\plund$ for large $\l$s. Note that the distribution
of the data in the ensemble (i.e., the likelihood in the integrand
above) is given by the true likelihood $\mathcal{L}_\Gamma$, not
the one used in the estimation.

Since the mean of the posterior distribution, $\plbar$, does not
necessarily coincide with $\plhat$ except for
$\mathcal{P}_\mathrm{G,cc}$, it may be less ideal
as an estimator. For instance, $\mathcal{P}_\Gamma$ leads to
\begin{align}
\plbar &= \int \pl \mathcal{P}_{\Gamma}(\pl | \plhat) \d\pl =
\frac{2\l + 1}{2\l - 3} \plhat \\
\langle \plbar \rangle &= \int \plbar
\mathcal{L}_{\Gamma}(\plhat|\plund) \d\plhat =
\frac{2\l + 1}{2\l - 3} \plund \\
\sigma_{\plbar}^2 &= \int \plbar^2
\mathcal{L}_{\Gamma}(\plhat|\plund) \d\plhat -
\langle \plbar \rangle^2 =
\frac{(4\l + 2)\left(\plund\right)^2}{(2\l - 3)^2}
\end{align}
for $\l \ge 2$, which perform poorly at low $\l$s.
Moreover, Equations~(\ref{eq:L_G}) and (\ref{eq:L_Gnd})
show that as $\pl$ approaches infinity, the posterior
$\mathcal{P}_\mathrm{G}$ behaves as $\pl^{-1}$,
and $\mathcal{P}_\mathrm{G,nd}$ becomes a constant
of $\pl$. Consequently, all the moments of $\pl$ diverge with
$\mathcal{P}_\mathrm{G}$ and $\mathcal{P}_\mathrm{G,nd}$.
Even though $\mathcal{P}_\mathrm{G}$ appears to underestimate the
uncertainty of the quadrupole in Figure~\ref{fig:like-post}, all
multipoles suffer from an infinite mean value and infinite
variance with $\mathcal{P}_\mathrm{G}$ and $\mathcal{P}_\mathrm{G,nd}$.
As such, a prior should be applied to limit the parameter range
when using $\mathcal{L}_\mathrm{G}$ and $\mathcal{L}_\mathrm{G,nd}$
to estimate parameters that can drive $\pl$ to
infinity (e.g., the normalization of the power spectrum).

What about parameters that are generic functions of $\pl$? In this
case, one usually obtains the posterior of the parameter via
 $\mathcal{P}(\theta | \plhat) \propto P(\theta)
\mathcal{L}(\plhat | \theta) \equiv P(\theta)
\mathcal{L}[\plhat | \pl(\theta)]$, where $P(\theta)$ is the
prior on $\theta$.
Taking $\theta=\ln\pl$ as an example and applying a flat prior on
$\theta$, one gets a biased estimator using the mode
of $\mathcal{P}(\theta | \plhat)$,
\begin{align}  \nonumber
\langle \tilde{\theta} \rangle &= \int \tilde{\theta}\,
\mathcal{L}_{\Gamma}(\plhat|\plund) \d\plhat =
 \int \ln \plhat\,
\mathcal{L}_{\Gamma}(\plhat|\plund) \d\plhat \\ \label{eq:theta}
&= \psi \left(\frac{2\l + 1}{2}\right) +
\ln \left(\frac{2}{2\l + 1}\right) + \theta^\mathrm{u},
\end{align}
where $\psi$ is the digamma function, and
$\theta^\mathrm{u} \equiv \ln \plund$. The difference
$\langle\tilde{\theta}\rangle - \theta^\mathrm{u}$ scales as
$-(2\l)^{-1}$ for large $\l$s. One could also map the posterior of
$\theta$ from that of $\pl$ via $\mathcal{P}(\theta | \plhat) =
\mathcal{P}(\pl | \plhat) \d\pl/\d\theta$, which
is equivalent to applying
$P(\theta) = P(\pl) \d\pl/\d\theta$
with the conventional approach. The result would still be biased.

Maximum likelihood estimators are asymptotically unbiased, meaning
that the biases, if exist, decrease with the sample size. This is
indeed seen in Equation~(\ref{eq:theta}): the bias of $\theta$ 
estimated from a particular $\pl$ is inversely proportional to the 
number of modes available at $\l$. The asymptotic behavior seems to
guarantee unbiased parameter estimation with abundant 
small-scale data. However, some parameters might not be sensitive to 
small scales at all, and the rate at which the bias drops with the 
sample size may also vary with the parameters. Given that much work 
has been done to improve approximations of the power spectrum 
likelihood function, it is necessary as well to quantify potential 
biases on the parameters estimated with even the most accurate power 
spectrum likelihood.

Determining the uncertainties is an integral part of parameter 
estimation. For an observed quadrupole $\hat{P}_2$ in a full-sky
noise-free survey, the 1-$\sigma$ 
confidence interval of the estimated $P_2$ from the exact posterior 
$\mathcal{P}_\Gamma$, i.e., the range enclosing 68\%  posterior 
probability and having the same probability density at its two end 
points, is $[0.42\hat{P}_2, 3.41\hat{P}_2]$ (see the upper-right 
panel of Figure~\ref{fig:like-post}). Given the \emph{WMAP} 9-year 
maximum likelihood value of $151 \mu K^2$ for the quadrupole
\citep{bennett13}, the 1-$\sigma$ interval is then 
$[63.4\mu K^2, 515\mu K^2]$ in the best case, which is slightly 
narrower than that in \citet{bennett13}. 

Even with the ideal survey considered above, the uncertainties 
of the power spectrum inferred from its posterior can differ 
considerably from the cosmic variance on scales where the number 
of modes available is small. For instance, the ensemble averaged 
1-$\sigma$ confidence interval of the quadrupole, i.e.,
$[0.42P_2^\mathrm{u}, 3.41P_2^\mathrm{u}]$, 
is more than twice the range of cosmic variance, i.e., 
$[0.37P_2^\mathrm{u}, 1.63P_2^\mathrm{u}]$.
When the scales of interest are much smaller than the survey 
dimensions, the uncertainties become essentially the same as the 
cosmic variance (in practice, noise dominates the small-scale 
uncertainties).

\subsection{Estimation with Fiducial Data}
\label{sec:fiducial}

To study the effect of the power spectrum likelihood in a more
realistic way, we employ a Markov Chain Monte Carlo (MCMC)
code, {\sc CosmoMC}\footnote{http://cosmologist.info/cosmomc/}
\citep{lewis02}, to estimate parameters from mock galaxy angular
power spectra and compare the results.
Because $\mathcal{L}_\mathrm{G,nd}$ is not widely used, hereafter
we only consider $\mathcal{L}_{\Gamma}$, $\mathcal{L}_\mathrm{G}$,
$\mathcal{L}_\mathrm{G,cc}$ and $\mathcal{L}_\mathrm{G+LN}$.

We assume an imaging survey covering half of the sky and a Gaussian
galaxy redshift distribution centered at $z_\mathrm{m}=1$ with a
dispersion of $\sigma_z=0.15$. The surface number density is taken to
be $\ng=10\,\arcm^{-2}$. The calculation of the angular power
spectrum is described in \citet{zhan06} with modification to include
the damping of the BAO signal \citep{eisenstein07}. Since the
difference between various approximate likelihoods and the exact one
is most pronounced on the largest scales, we limit the multiple
range to $2 \le \l \le 1000$.

The parameters include the matter fraction $\Om$, tilt of
the matter power spectrum $\ns$, reduced Hubble constant $h$,
rms density fluctuation within $8\,\hmpc$ $\s8$, primordial
non-Gaussianity parameter of the local type $\fnl$, and linear
clustering bias $\bg$. The fiducial values of these parameters are
$(\Om, \ns, h, \s8, \fnl, \bg) = (0.27, 0.96, 0.72, 0.78, 0, 2)$,
consistent with recent measurements \citep{larson11}.
The baryon fraction is fixed at $\Ob = 0.0446$ and the cosmological
constant parameter $\Omega_\Lambda=1-\Om$.

The primordial non-Gaussianity is one of the few probes that can
help shed light on the physics of inflation. Therefore, the
parameter $\fnl$ is of great interest to future galaxy surveys.
It gives rise to a
scale-dependent effective bias \citep{dalal08, matarrese08}
\be
\label{eq:b}
b(k,\fnl) = \bg + \fnl (\bg - 1) u(k,z),
\ee
where $u(k,z)\propto k^{-2} T^{-1}(k)$ with $T(k)$ being the transfer
function of fluctuations \citep[see e.g.,][]{hamaus11}. 
Because the effect of $\fnl$ is most pronounced on the largest
scales, its estimation may be prone
to the errors of the approximate likelihoods at low multipoles.

In this subsection, we set the ``observed'' power spectrum $\plhat$ 
to the power spectrum of the fiducial cosmological model, i.e., 
$\plund$, which in some sense represents the best case scenario. The 
galaxy angular power spectrum with the effective bias can be written 
as 
\be
\label{eq:pl-fnl}
\pl = \plund + \fnl Q_{\l}+\fnl^2 R_{\l},
\ee
where $Q_{\l}$ and $R_{\l}$, like $\plund$, are determined by 
parameters other than $\fnl$. Equation~(\ref{eq:pl-fnl}) is 
convenient when $\fnl$ is the only parameter to be estimated.

\begin{figure}
\includegraphics[width=\columnwidth]{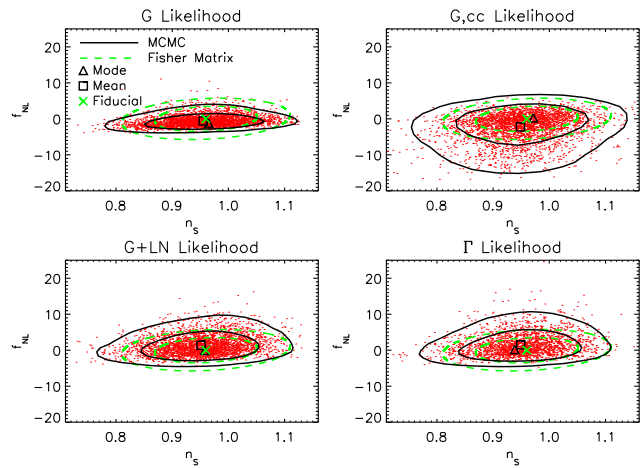}
 \caption{Constraints on the primordial non-Gaussianity parameter
$\fnl$ and the power spectrum tilt $\ns$. Each panel
presents the results derived from the likelihood of the galaxy
angular power spectrum as labeled. The black solid contours show
the 1-$\sigma$ and 2-$\sigma$ regions of the MCMC samples
(red dots), and
the green dashed ones denote Fisher matrix forecasts. The fiducial,
mode, and mean values of the parameters from MCMC samples are marked
with crosses, open triangles, and open squares, respectively.}
\label{fig:mcmc-6p}
\end{figure}

We first estimate all 6 parameters simultaneously. A Gaussian 
prior is applied to the linear
bias $\sigma(\bg)/\bg=0.2$, and flat priors are applied in the
following parameter ranges: $0<\Om<1$, $0.5<\ns<1.5$, $0<h<2$,
$0<\s8<2$, and $-200<\fnl<200$.
For each of the four likelihoods, we run four chains with about
$10^5$ samples per chain after burn-in and merge them into one
sample. The marginalized constraints on $\ns$ and $\fnl$ are shown
in Figure~\ref{fig:mcmc-6p}. The black solid contours indicate the
1-$\sigma$ and 2-$\sigma$ regions estimated from the MCMC samples
(red dots, thinned to leave $\sim3000$ points for display), with
the mode and mean values marked by triangles and squares,
respectively. The green dashed contours denote predictions by
a Fisher matrix analysis centered at the fiducial values
(green crosses), which is  based on the likelihood form of 
$\mathcal{L}_\mathrm{G,cc}$ \citep{carron13}.

When all the other parameters are marginalized, all the four
likelihoods are able to recover the fiducial values of $\ns$ and
$\fnl$ without much bias. However, the shape of the posterior
contours based on $\mathcal{L}_\mathrm{G}$ and
$\mathcal{L}_\mathrm{G,cc}$ differ significantly from that based on
the correct likelihood $\mathcal{L}_{\Gamma}$, leading to 
mis-estimation of the uncertainties of $\fnl$. Contours given by 
$\mathcal{L}_\mathrm{G+LN}$, meanwhile, agree with those given by 
$\mathcal{L}_{\Gamma}$.

Equation~(\ref{eq:b}) means that a positive $\fnl$ would increase 
the power spectrum as long as $\bg>1$. Since the posterior 
probabilities of low multipoles based on $\mathcal{L}_\mathrm{G}$ and
$\mathcal{L}_\mathrm{G,cc}$ are significantly lower than those based 
on $\mathcal{L}_\mathrm{G+LN}$ and $\mathcal{L}_{\Gamma}$ at 
$\pl > \plhat$ (e.g., Figure~\ref{fig:like-post}), the $\fnl$-$\ns$ 
contours in the $\mathcal{L}_\mathrm{G}$ and $\mathcal{L}_\mathrm{G,cc}$
panels of Figure~\ref{fig:mcmc-6p} are less extended toward the 
positive $\fnl$ direction than those in the 
$\mathcal{L}_\mathrm{G+LN}$ and $\mathcal{L}_{\Gamma}$ panels.
Similarly, the contours in the $\mathcal{L}_\mathrm{G,cc}$ panel is 
more extended in the negative $\fnl$ direction than those in the 
other panels. 

The effective bias could be driven below zero numerically by $\fnl$ 
so that one might expect a second peak in the posterior of $\fnl$. 
This is actually true for a single mode in $k$ 
($\sim \l/D_\mathrm{A}$ with $D_\mathrm{A}$ being the angular diameter 
distance). When all the modes are included, the probability of a 
large negative $\fnl$ is strongly suppressed by low multipoles.
As a related test, we impose a cut $b>0$ in MCMC. The result of 
$\mathcal{L}_\mathrm{G,cc}$ becomes much more consistent with that 
of the Fisher Matrix analysis, and there is essentially no change 
in the results of the other likelihoods.

\begin{figure}
\includegraphics[width=\columnwidth]{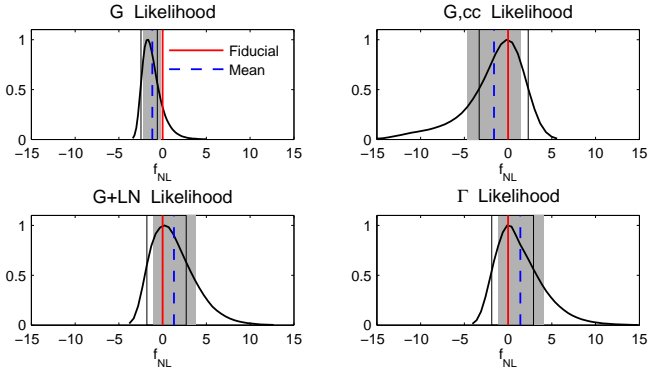}
 \caption{Posterior probability distribution of the primordial
non-Gaussianity parameter $\fnl$ with all the other parameters fixed.
Each panel presents the results derived from the likelihood of the
galaxy angular power spectrum as labeled. Black solid curves show the
probability distributions of $\fnl$ with the minimum confidence 
intervals enclosed by thin vertical lines. The mean values and 
corresponding 1-$\sigma$ central confidence intervals are denoted by 
dashed lines and grey shaded regions, respectively.
The red solid lines mark the fiducial values of $\fnl$.}
 \label{fig:scan-fnl}
\end{figure}

Next, we fix all the other parameters except $\fnl$. In this case, 
the posterior can be scanned efficiently using 
Equation~(\ref{eq:pl-fnl}). The scan range is $[-200,200]$ with a 
step size of $0.01$. The results are shown in 
Figure~\ref{fig:scan-fnl}. The Gaussian likelihood 
$\mathcal{L}_\mathrm{G}$ causes a significant bias on $\fnl$ and 
underestimates the uncertainty. The Gaussian 
with a constant covariance likelihood $\mathcal{L}_\mathrm{G,cc}$ 
does not bias $\fnl$ much but causes a spurious tail on the 
negative side. The Gaussian plus log-normal likelihood 
$\mathcal{L}_\mathrm{G+LN}$ again reproduces the result of the 
correct likelihood $\mathcal{L}_{\Gamma}$ with less than $10\%$ 
errors in the mean value and the two types of confidence level 
intervals in Figure~\ref{fig:scan-fnl}.

\subsection{Ensemble Behavior of the Estimators}
\label{sec:real}

We generate $10^4$ sets of galaxy angular power spectra to examine 
the ensemble behavior of the estimators of the primordial 
non-Gaussianity parameter. The power of each multipole 
($2\le \l \le 1000$) is randomly drawn from its underlying Gamma 
distribution. We then scan the posterior probability of $\fnl$ from 
these power spectra using the same likelihood functions as in the 
last subsection. The scan range and step size are also kept the same.

\begin{figure}
\includegraphics[width=\columnwidth]{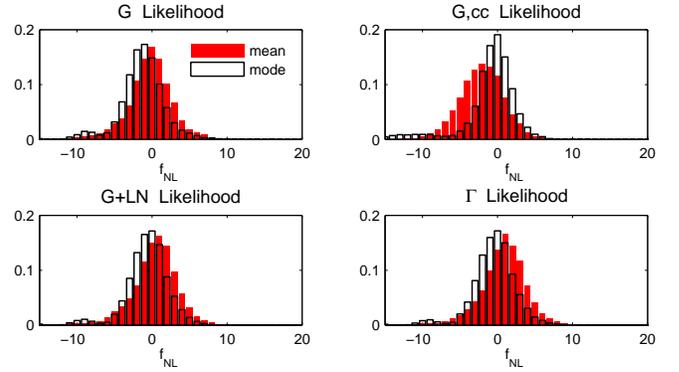}
\caption{Ensemble distributions of the mean $\fnl$ (filled bars) and
the maximum likelihood $\fnl$ (open bars) estimated from $10^4$
realizations of the ``observed'' power spectrum.
Each panel presents the results from the likelihood of the galaxy
angular power spectrum as labeled.}
 \label{fig:hist}
\end{figure}

The mean value $\bar{f}_\mathrm{NL}$ and the maximum likelihood
value $\tilde{f}_\mathrm{NL}$ are determined for each set of the 
galaxy angular power spectrum. Their ensemble distributions are 
shown in Figure~\ref{fig:hist}. One can see minor peaks of 
$\tilde{f}_\mathrm{NL}$ near $-10$. This is caused by the same
negative effective bias issue noted in the last subsection where 
the posterior of a very special realization is examined.
These minor peaks disappear when a cut of $b>0$ is imposed. 

The distribution of $\bar{f}_\mathrm{NL}$ from 
$\mathcal{L}_\mathrm{G,cc}$ is mildly skewed toward $\fnl < 0$,
and all the other results display no significant bias. The
maximum likelihood estimator of $\mathcal{L}_\mathrm{G,cc}$,
$\mathcal{L}_\mathrm{G+LN}$ and $\mathcal{L}_\Gamma$ are not biased,
if the minor peaks are ignored. However, it is worth mentioning that
the ensemble behavior is different from the behavior of a particular
realization. As the upper left panel of Figure~\ref{fig:scan-fnl} 
shows, the Gaussian likelihood $\mathcal{L}_\mathrm{G}$ underestimates 
$\fnl$ by more than 1-$\sigma$ even with $\plhat$ set to $\plund$.

\section{Summary}
\label{sec:sum}

We have examined several approximate likelihood functions for 
parameter estimation with the galaxy angular power spectrum in 
an idealized case. 
None of the three forms of Gaussian likelihoods in the study 
(Gaussian $\mathcal{L}_\mathrm{G}$, Gaussian without the 
determinant $\mathcal{L}_\mathrm{G,nd}$, and Gaussian with 
a constant covariance $\mathcal{L}_\mathrm{G,cc}$) can accurately 
reproduce the posterior probability of the power spectrum on scales 
where the number of modes contained in the survey is small. 
This issue exists regardless of the physical dimensions of the 
survey because the distribution of the power spectrum is 
intrinsically not Gaussian. The distribution approaches Gaussian 
only on scales much smaller than the survey size where 
there are enough modes for the central limit theroem 
to take effect. Applying the three Gaussian approximations 
may lead to biases on parameters that are 
constrained mainly by scales close to the survey size.

In our tests with the primordial non-Gaussianity parameter $\fnl$, 
the Gaussian likelihoods $\mathcal{L}_\mathrm{G}$ and 
$\mathcal{L}_\mathrm{G,cc}$ lead to distorted $\ns$-$\fnl$ error 
contours when four other parameters are marginalized. A significant 
bias and underestimated uncertainty on $\fnl$ are obtained with
$\mathcal{L}_\mathrm{G}$ when all other parameters are fixed. The
Gaussian plus log-normal likelihood $\mathcal{L}_\mathrm{G+LN}$ 
can reproduce the true posterior probability of the power spectrum 
and that of $\fnl$ accurately. 

Analyses with real data often have to work with far more complex 
likelihood functions than we have considered. It is also worth 
emphasizing that 
using the exact likelihood of the power spectrum does not guarantee 
unbiased estimates of the parameters. 
Although perfection of the likelihood approximations has been
pursued (mostly in CMB analyses), relatively less effort has been 
made on quantifying potential biases of the maximum likelihood 
estimator itself,
which would be specific to each unique set of parameters and data. 
Given the unprecedented statistical power of future 
surveys, it is especially important to thoroughly examine the 
likelihood functions and estimators for potential 
systematics. 

\acknowledgments
We thank the referee for the very constructive comments and suggestions.
This work was supported by the National Natural Science Foundation of
China grant No. 11033005, the National Key Basic Research Science
Foundation of China grant No. 2010CB833000, the Bairen program from
the Chinese Academy of Sciences.

%\bibliographystyle{apj}
%\bibliography{ms}

\end{document}